\documentclass{article}
\usepackage{arxiv}
\usepackage[utf8]{inputenc} 
\usepackage[T1]{fontenc}    
\usepackage{url}            
\usepackage{booktabs}       
\usepackage{amsfonts}       
\usepackage{nicefrac}       
\usepackage{microtype}      
\usepackage{graphicx}
\usepackage{epstopdf, epsfig}
\usepackage{subfigure}
\usepackage{amsmath,amssymb}
\usepackage{xargs}
\usepackage[pdftex,dvipsnames]{xcolor}  
\usepackage[textwidth=3.7cm,colorinlistoftodos,textsize=tiny]{todonotes}
\newcommandx{\unsure}[2][1=]{\todo[linecolor=red,backgroundcolor=red!25,bordercolor=red,#1]{#2}}
\newcommandx{\change}[2][1=]{\todo[linecolor=blue,backgroundcolor=blue!25,bordercolor=blue,#1]{#2}}
\newcommandx{\info}[2][1=]{\todo[linecolor=OliveGreen,backgroundcolor=OliveGreen!25,bordercolor=OliveGreen,#1]{#2}}
\newcommandx{\improvement}[2][1=]{\todo[linecolor=Plum,backgroundcolor=Plum!25,bordercolor=Plum,#1]{#2}}
\newcommandx{\addref}[2][1=]{\todo[linecolor=green,backgroundcolor=green!25,bordercolor=red,#1]{need ref: #2}}
\newcommandx{\thiswillnotshow}[2][1=]{\todo[disable,#1]{#2}}
\title{Evaluating the leeway coefficient for different ocean drifters using operational models}
\author{
  Graig Sutherland\\
  Environmental Numerical Prediction Research\\
  Environment and Climate Change Canada\\
  Dorval, QC\\
   \And
  Nancy Soontiens\\
  North Atlantic Fisheries Center\\
  Fisheries and Oceans Canada\\
  St. John's, NL, Canada\\
   \And
  Fraser Davidson\\
  North Atlantic Fisheries Ceter\\
  Fisheries and Oceans Canada\\
  St. John's, NL, Canada\\
   \And
  Gregory C. Smith\\
  Environmental Numerical Prediction Research\\
  Environment and Climate Change Canada\\
  Dorval, QC, Canada\\
   \And
  Natacha Bernier\\
  Environmental Numerical Prediction Research\\
  Environment and Climate Change Canada\\
  Dorval, QC, Canada\\
   \And
  Hauke Blanken\\
  Institute of Ocean Sciences\\
  Fisheries and Oceans Canada\\
  Sidney, BC, Canada\\
  \And
  Douglas Schillinger\\
  Beford Institute of Oceanography\\
  Fisheries and Oceans Canada\\
  Dartmouth, NS, Canada\\
  \And
  Guillaume Marcotte\\
  Environmental Emergency Response Section\\
  Environment and Climate Change Canada\\
  Dorval, QC, Canada\\
  \And
  Johannes R{\"o}hrs\\
  Norwegian Meteorological Institute\\
  Oslo, Norway
   \And
  Knut-Frode Dagestad\\
  Norwegian Meteorological Institute\\
  Bergen, Norway
   \And
  Kai H. Christensen\\
  Norwegian Meteorological Institute\\
  Oslo, Norway
   \And
  {\O}yvind Breivik\\
  Norwegian Meteorological Institute\\
  Bergen, Norway
}

\begin{document}
\maketitle

\begin{abstract}

   The water following characteristics of six different drifter types are investigated using two different operational marine environmental prediction systems: one produced by Environment and Climate Change Canada (ECCC) and the other produced by the Norwegian Meteorological Institute (METNO). These marine prediction systems include ocean circulation models, atmospheric models, and surface wave models. Two leeway models are tested for use in drift object prediction: an implicit leeway model where the Stokes drift is implicit in the leeway coefficient, and an explicit leeway model where the Stokes drift is provided by the wave model. Both leeway coefficients are allowed to vary in direction and time in order to perfectly reproduce the observed drifter trajectory. This creates a time series of the leeway coefficients which exactly reproduce the observed drifter trajectories.  Mean values for the leeway coefficients are consistent with previous studies which utilized direct observations of the leeway. For all drifters and models, the largest source of variance in the leeway coefficient occurs at the inertial frequency and the evidence suggests it is related to uncertainties in the ocean inertial currents.

\end{abstract}



\section{Introduction}
\label{sec:introduction}

Accurate knowledge of surface currents are important for predicting the transport of buoyant material in the ocean~\citep{Christensen_etal_2018}. Examples of typical material in the ocean are oil spills~\citep{Spaulding_2017}, marine debris~\citep{vanSebille_etal_2015, Jansen_etal_2016}, as well as natural occurring material related to biology such as fish eggs and larvae~\citep{Sundby_1983}. In general, the modeling of material transport assumes that the only dynamic physical property of the material is it's buoyancy, and the horizontal motion is that of a passive tracer. However, for objects at the surface it is common to add direct wind forcing on the object, commonly referred to as "leeway", but also the terms "windage" or "wind-slip" are sometimes used~\citep{Niiler_etal_1995}, in addition to the ocean currents. The windage can be derived from the ratio of air-side drag to the ocean-side drag~\citep{Kirwan_etal_1975}, but in practice it also compensates for effects due to finite vertical resolution in the ocean model~\citep{Isern-Fontanet_etal_2017,Tamtare_etal_2019}, missing physics such as the Stokes drift~\citep{vandenBremer_Breivik_2017}, or uncertainties in the operational prediction system~\citep{Dagestad_Rohrs_2019}. The horizontal motion will also vary with depth due to the rapid decay of the Stokes drift with increasing depth~\citep{Breivik_etal_2014} and unresolved shear near the ocean surface~\citep{Laxague_etal_2018}. 

In general, the leeway is often parameterized as a linear function of the wind speed~\citep{Breivik_etal_2011}. While the leeway coefficient has been shown to be equivalent to the ratio of the air-side drag to the water-side drag~\citep{Kirwan_etal_1975}, it is far from trivial on how to estimate the drag ratio for an object bobbing at the ocean surface. This is especially true for more exotic objects than spherical drifters~\citep{Breivik_etal_2012}. Therefore, it was suggested by~\citet{allen1999} and~\citet{Breivik_etal_2011} that the leeway coefficient should be estimated using a direct method where detailed observations of the relative velocity between the drifter and the ocean current are compared with the wind velocity. Experiments exist for estimating the leeway coefficient using the direct method for some ocean drifters~\citep{Niiler_etal_1995,Poulain_Gerin_2019}, as well as a wide range of objects encountered in search and rescue~\citep{Allen_2005,Breivik_etal_2011}. However, these experiments are relatively rare, expensive and rely on the universality of the calculated leeway, i.e. a drift object taxonomy~\citep{allen1999,Allen_2005,breivik08operational}, which is applicable to a class of drifting objects~\citep{Breivik_etal_2011}. In addition, the direct method does not directly assess the ability to predict the drift trajectory using operational marine prediction systems, which will include effects due to finite resolution and parameterized physics. 

So what is the best method for determining an optimal leeway coefficient to use in an operational prediction system? The leading method used to date is to calculate a skill score based on the observed and modeled trajectories using a range of values for the leeway coefficients~\citep{Toner_etal_2001,Molcard_etal_2009,Liu_Weisberg_2011,Rohrs_etal_2012,Dagestad_Rohrs_2019}. However, the skill scores in these studies are all based on the separation distance, or by the time-averaged separation distance as in the score by~\citet{Liu_Weisberg_2011}, between the two trajectories, which implies that they are sensitive to the timing of the errors as large errors early in the trajectory will have a tendency to accumulate over time, especially in regions with appreciable horizontal shear. In addition, most of these studies restrict the leeway coefficient to a scalar~\citep{Rohrs_etal_2012, Dagestad_Rohrs_2019} and may struggle in regions with large uncertainties in oscillating currents, e.g. inertial oscillations, which can have a significant impact on drift prediction at time scales on the order of the inertial period~\citep{Christensen_etal_2018}. 

In this paper we take a new approach to estimating the leeway coefficient. In this approach, the leeway coefficient is allowed to vary in magnitude, direction and time in order to reproduce an exact trajectory given a particular input of ocean, wind and wave fields from two operational prediction systems. The model velocities are interpolated to the drifter positions in time and the leeway coefficient can be estimated directly. This has two large advantages to using forecast skill scores: one, that all the velocity values are equally weighted in time, and two, that the maximum of the leeway probability distribution function will provide the best estimate assuming the forcing models are unbiased. As this method allows the leeway coefficients to vary in time in order to give a perfect trajectory, it also provides statistics which can be associated with uncertainties in the prediction systems. The results are presented using six different types of drifters and two operational prediction systems for the ocean current, wind and Stokes drift; one from Environment and Climate Change Canada (ECCC) and the other from MET Norway (METNO). The outline is as follows. Section~\ref{sec:leeway} presents the leeway models used in this study. Details about the drifters and the two operational prediction systems are found in section~\ref{sec:data}. Results are presented in section~\ref{sec:results} followed by a discussion in section~\ref{sec:discussion} and a summary of the results in section~\ref{sec:conclusions}.

\section{Leeway model}
\label{sec:leeway}

The standard leeway model~\citep{allen1999,Allen_2005} as implemented by~\citet{breivik08operational} and more recently by~\citet{dagestad18} has all the effects due to finite resolution and missing physics, including the Stokes drift, implicit in the leeway coefficient and is given by the equation
\begin{equation}\label{eq:eulerian_leeway}
	\mathbf{u}_d = \mathbf{u}_o + \alpha\mathbf{U}_{10},
\end{equation}
where $\mathbf{u}_d$ is the drift velocity vector, $\mathbf{u}_o$ is the ocean velocity vector at the effective depth of the drifter, $\mathbf{U}_{10}$ is the 10 m wind speed vector and $\alpha$ is the leeway coefficient. Henceforth, (\ref{eq:eulerian_leeway}) will be referred to as the implicit leeway model and $\alpha$ is the implicit leeway coefficient. The coefficient $\alpha$ is commonly a scalar, implying the drag is only in the along-wind direction, but it can also be a vector as some objects at sea have implicit leeway coefficients with both downwind and cross-wind components~\citep{Allen_2005}. The implicit leeway coefficient is used to parameterize a broad range of processes from a direct wind drag on the drifter~\citep{Niiler_etal_1995}, to compensate for missing physics due to feedback on the forcing fields or compensate for inadequate resolution, and most prominently the Stokes drift, which is the Lagrangian drift due to the surface waves. For a fully-developed sea, the Stokes drift at the surface is typically about 1.0 to 1.5\% of $\mathbf{U}_{10}$ and decays rapidly with depth~\citep{Breivik_etal_2014,Breivik_etal_2016}. 

As the Stokes drift is increasingly available from operational wave prediction systems, it can be included explicitly in the trajectory model. We will define an explicit leeway model, which explicitly includes the contribution from the Stokes drift, as
\begin{equation}\label{eq:lagrangian_leeway}
   \mathbf{u}_d = \mathbf{u}_o + \mathbf{u}_s + \beta\mathbf{U}_{10},
\end{equation}
where $u_{s}$ is the Stokes drift at the effective depth of the drifter and $\beta$ is the leeway coefficient when the Stokes drift is explicitly included. We will refer to this model as the explicit leeway model. As mentioned in the previous paragraph, the Stokes drift is reasonably estimated by using a percentage of the wind velocity and is often not explicitly included in leeway modelling~\citep{Breivik_etal_2011}. In general, the depth-dependent Stokes drift is not a standard output variable of ocean wave prediction systems, but can be estimated from the more common surface Stokes drift ($\mathbf{u}_{s0}$) and assuming the exponential decay can be estimated by a single wavenumber $\bar{k}$, i.e.~\citep{Breivik_etal_2014}
\begin{equation}\label{eq:stokes}
   \mathbf{u}_s = \mathbf{u}_{s0} e^{2\bar{k}z},
\end{equation}
where $z$ is the depth (negative from surface) and $\bar{k}$ is the vertical wavenumber which can be estimated from the surface Stokes drift and the Stokes transport ($\mathbf{T}_{st}$)
\begin{equation}\label{eq:wavenumber}
   \bar{k} = \frac{\left|\mathbf{u}_{s0}\right|}{2\left|\mathbf{T}_{st}\right|},
\end{equation}
where the Stokes transport is defined as the vertical integral of the Stokes drift 
\begin{equation}\label{eq:stokestransport}
   \mathbf{T}_{st} = \int_{-\infty}^0 \mathbf{u}_s dz,
\end{equation}
and is a common output of ocean wave prediction systems.

Estimates of $\mathbf{u}_{s0}$ and $\mathbf{u}_{st}$ can also be made from the significant wave height ($H_S$: the mean height of the highest 1/3 waves) and the mean zero-upcrossing period ($T_{z0}$: the mean period between successive times when the water elevation crosses from below to above the mean elevation) if only these estimates of the wave field are present. The magnitude of the surface Stokes drift and the Stokes transport can be estimated to be
\begin{equation}\label{eq:stokes2}
   u_{s0} = \frac{\pi^3 H_S^2}{gT_{z0}^3},
\end{equation}
and
\begin{equation}\label{eq:transport2}
   T_{st} = \frac{2\pi H_S^2}{16 T_{z0}}
\end{equation}
respectively. The direction is either given by a mean wave direction given by the wave model or assumed to be in the direction of the wind. It is important to note that (\ref{eq:stokes2}) and (\ref{eq:transport2}) are simplifications based on a deepwater dispersion relation of $\bar{k} = 4\pi^2/(gT_{z0}^2)$ and assuming the waves are predominantly unidirectional. The validity of using (\ref{eq:stokes2}) and (\ref{eq:transport2}) will depend on the frequency and directional distribution of the surface waves as well as the local depth and surface current which can impact the dispersion relation for surface gravity waves~\citep{kirby1989surface}.

\section{Data and Methods}
\label{sec:data}

\subsection{Drifters}
\label{sec:drifters}

\begin{figure}[ht]
   \noindent\includegraphics[width=\textwidth]{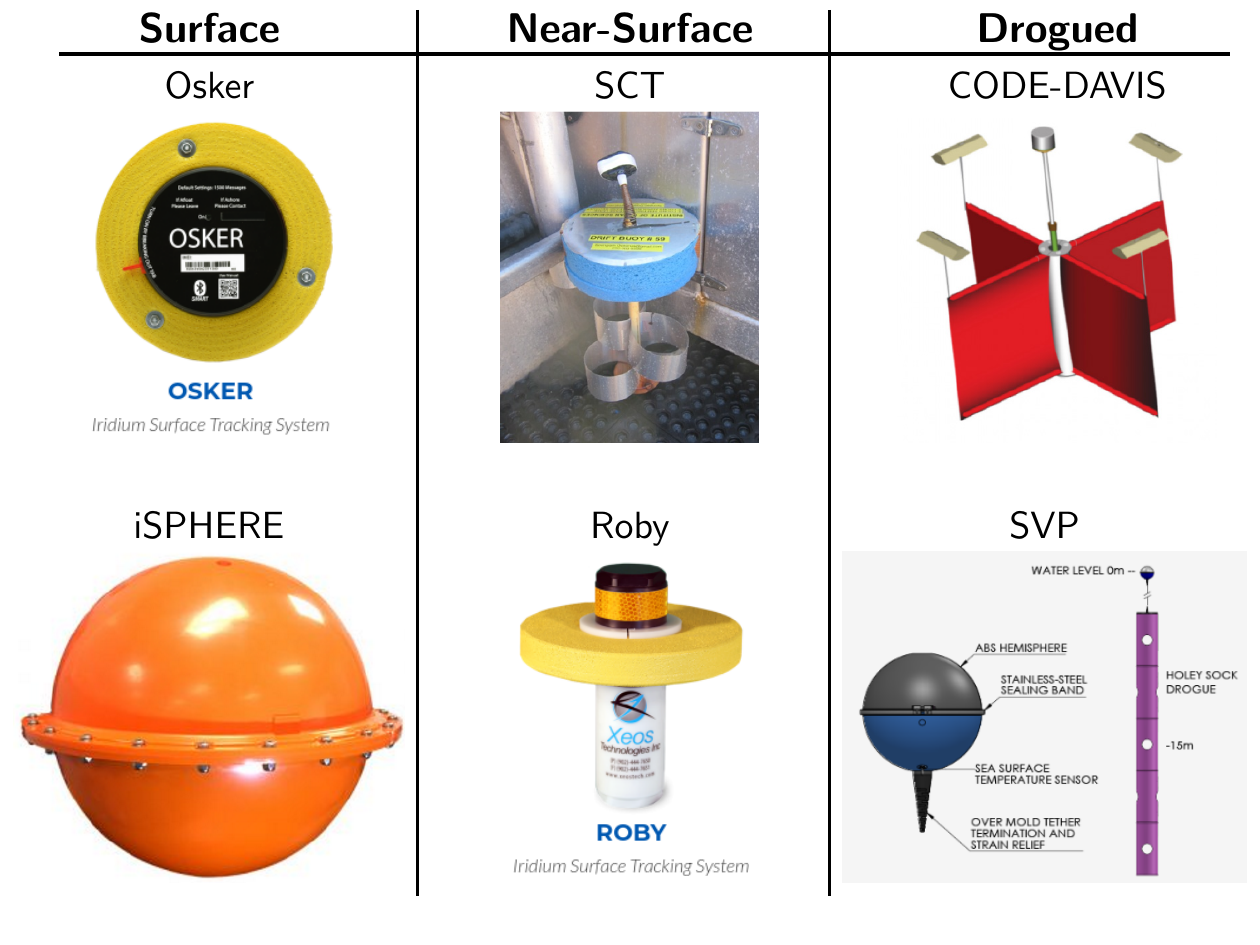}
   \caption{Drifters used in the experiment.}
   \label{fig:drifters}
\end{figure}

Six different types of drifters were deployed in this experiment, each representing a different effective depth with some having drogues to increase the ocean drag. These drifters are classified into three groups: surface, near-surface and drogued. This classification is predominantly a function of the wind response, with surface drifters having the largest wind effect, drogued having the least, and near-surface being somewhat intermediate. Therefore, these classifications are only loosely based on their effective depth and are more strongly related to the ratio of ocean-side drag to wind-side drag. An image of each drifter is shown in Figure~\ref{fig:drifters}.

We use two types of surface drifter, the disc shaped Osker drifter (Xeos Technologies Inc., Canada) and the spherical iSPHERE drifter (MetOcean, Canada). The Osker drifter is 12.7 cm in diameter and 5.1 cm height and is encased in a spherical foam ring of diameter 20.3 cm and height of 2 cm. The iSPHERE drifter is a slightly flattened sphere with a horizontal diameter of 39 cm and a vertical diameter of 31 cm. The iSPHERE drifters have been shown to follow the surface currents plus the surface Stokes drift~\citep{Rohrs_etal_2012} and are often used to represent the drift trajectories of oil~\citep{Beloire_etal_2011}. Therefore, we will also assume their effective depth to 0 cm.

In addition, there are two cylindrical drifters, which we will refer to as near-surface drifters, equipped with foam rings for flotation. These are the Roby drifter (Xeos Technologies Inc., Canada) and the Surface Circulation Tracker (SCT) (Oceanetic Measurement Ltd., Canada). The Roby drifter is 5.7 cm in diameter and 21 cm in length with a 20.3 cm diameter foam collar on the top for extra buoyancy. The SCT drifter is 50 cm in height with an expected draft of 33 cm and has a foam collar which is 23 cm in diameter and 7.5 cm thick. The effective depth of the Roby and SCT drifter is estimated to be 10 cm and 20 cm respectively.

The final two drifters use drogues and are the Surface Velocity Program (SVP) drifter~\citep{Niiler_etal_1995} and the Coastal Ocean Dynamics Experiment (CODE) drifter~\citep{Davis_1985}. The CODE drifter has a 0.9 m sail drogue centered at 0.6 m depth and the SVP drifter has a 6.1 m long holey-sock drogue with radius of 0.6 m and centered at 15 m depth. These mean depths of the drogues are taken to be the effective drift depth. While the CODE drifter drogue is in the upper meter, which for all intents and purposes is generally assumed to be the surface, due to it's design the wind slippage has been found to be about 0.1\% of the wind speed ~\citep{Poulain_Gerin_2019}, which is the same as the measured wind slippage for the SVP drifters~\citep{Lumpkin_2017}. Therefore, the direct wind forcing on the CODE and SVP drifters are expected to be similar and differences will be due to vertical shear and the Stokes drift.

\begin{figure}[ht]
	\noindent\includegraphics[width=\textwidth]{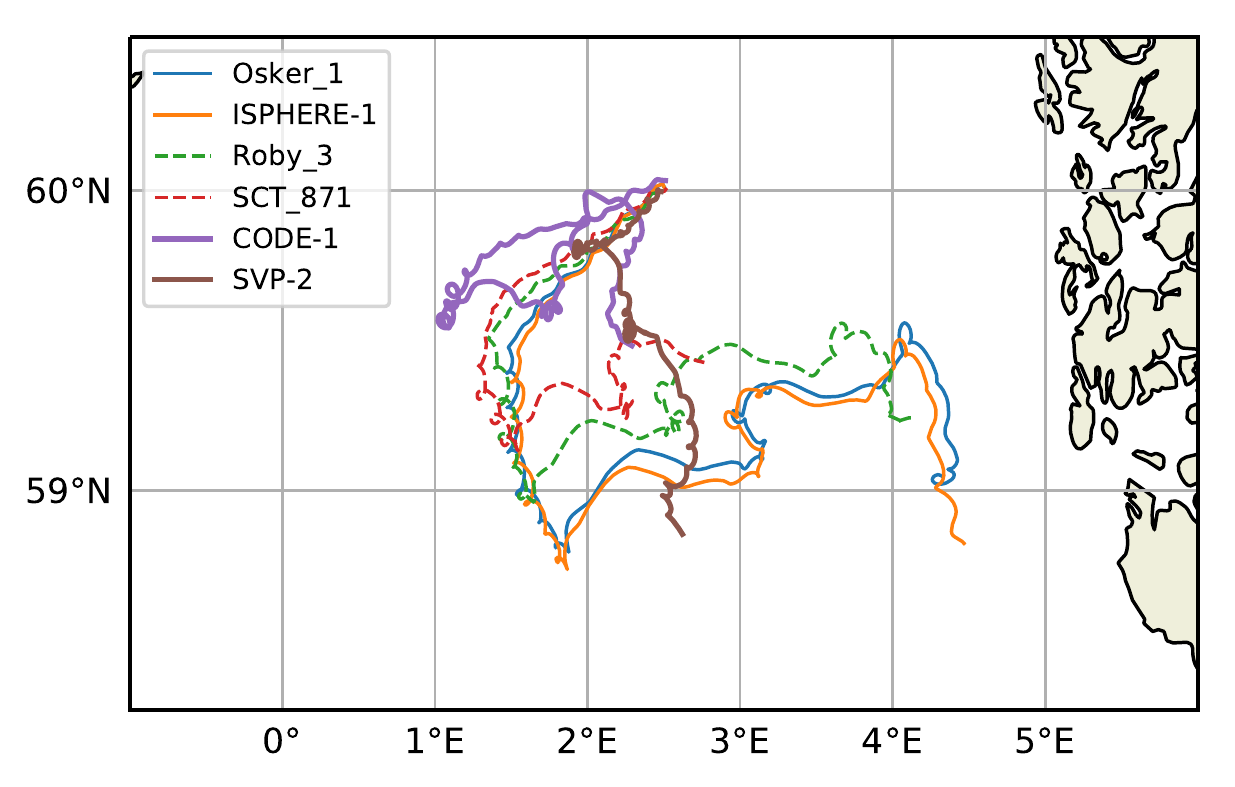}
	\caption{Trajectories for the various drifters over the period 6-18 June 2018. Thin solid lines are surface drifters, dashed lines are near-surface drifters and thick solid lines are the drogued drifters.}
   \label{fig:map}
\end{figure}

The six drifter types were deployed on 6 June 2018 and all were operational until 18 June 2018. After this time some of the drifters stopped sending their positions. Therefore, the analysis will focus on these 12 days when all the drifters were operational. An overview of the six drifter tracks is shown in Figure~\ref{fig:map}. The sampling rate for the drifters varied between 5 minutes and 1 hour. The velocity is calculated via forward difference of successive locations and the velocities are interpolated to hourly output by taking hourly means.  

\subsection{Marine environmental prediction systems}
\label{sec:model}

The ECCC operational prediction system consists of the Regional Ice-Ocean Prediction System (RIOPS) for the ocean currents, the Canadian Arctic Prediction System (CAPS) for the atmospheric winds, and the Global Deterministic Wave Prediction System (GDWPS) for the wave model providing the Stokes drift. RIOPS is a regional model based on NEMO (Nucleus for European Modelling of the Ocean, http://www.nemo-ocean.eu) with a horizontal resolution of 1/12$^\circ$~\citep{Dupont_etal_2015}. Surface currents are output every 3 hours and the surface layer depth is 1 m. GDWPS is a global wave model based on Wavewatch III with a 1/4$^\circ$ horizontal resolution and is run twice a day~\citep{Bernier_etal_2016} 1/4$^\circ$. GDWPS resolves the wave spectrum with 36 logarithmically spaced frequencies between 0.035 to 1 Hz and 36 directions. Atmospheric model is the Canadian Arctic Prediction System (CAPS) which is a 3 km resolution model that covers the entire Arctic and some northern regions such as Norway. The dynamical core of CAPS is GEM (Global Environmental Multiscale), a non-hydrostatic model which solves the fully compressible Euler equations~\citep{Cote_etal_1998a, Cote_etal_1998b, Girard_etal_2014}, and is run operationally at ECCC. Both GEM and GDWPS output data at hourly resolution.

The METNO operational prediction system used in this study consists of the data-assimilative ocean model NorShelf~\citep{NORSHELF_METNO}, the spectral wave model WAM4~\citep{hasselmann88,WAM_METNO} and the control run from the MetCoOp Ensemble Prediction System (MEPS)~\citep{Muller_etal_2017, Bentsson_etal_2017, Frogner_etal_2019}. NorShelf is based on the Regional Ocean Modeling System~\citep{shchepetkin05} and nested into Topaz~\citep{Xie_etal_2017}, providing ocean currents with a horizontal resolution of 2.4km and a vertical resolution at the surface of about 0.5-1m. The wave model is a version of the MyWaveWAM model set up at a 4 km resolution with boundary conditions in the form of two-dimensional spectra from the operational ECMWF wave forecasts. The model resolves the wave spectrum with 36 logarithmically spaced frequencies from 0.0345 to 0.9702 Hz and 36 directions. The wind forcing for both the ocean and wave models are taken from the control member of MEPS, which has a native horizontal resolution of 2.5 km. These regional systems are forced or nested into the ECMWF global forecast system and hourly data are available from the Norwegian Meteorological Institute's thredds server [http://thredds.met.no]. Wind, waves and ocean surface values are available at hourly resolution.

\begin{figure}[ht]
	\noindent\includegraphics[width=\textwidth]{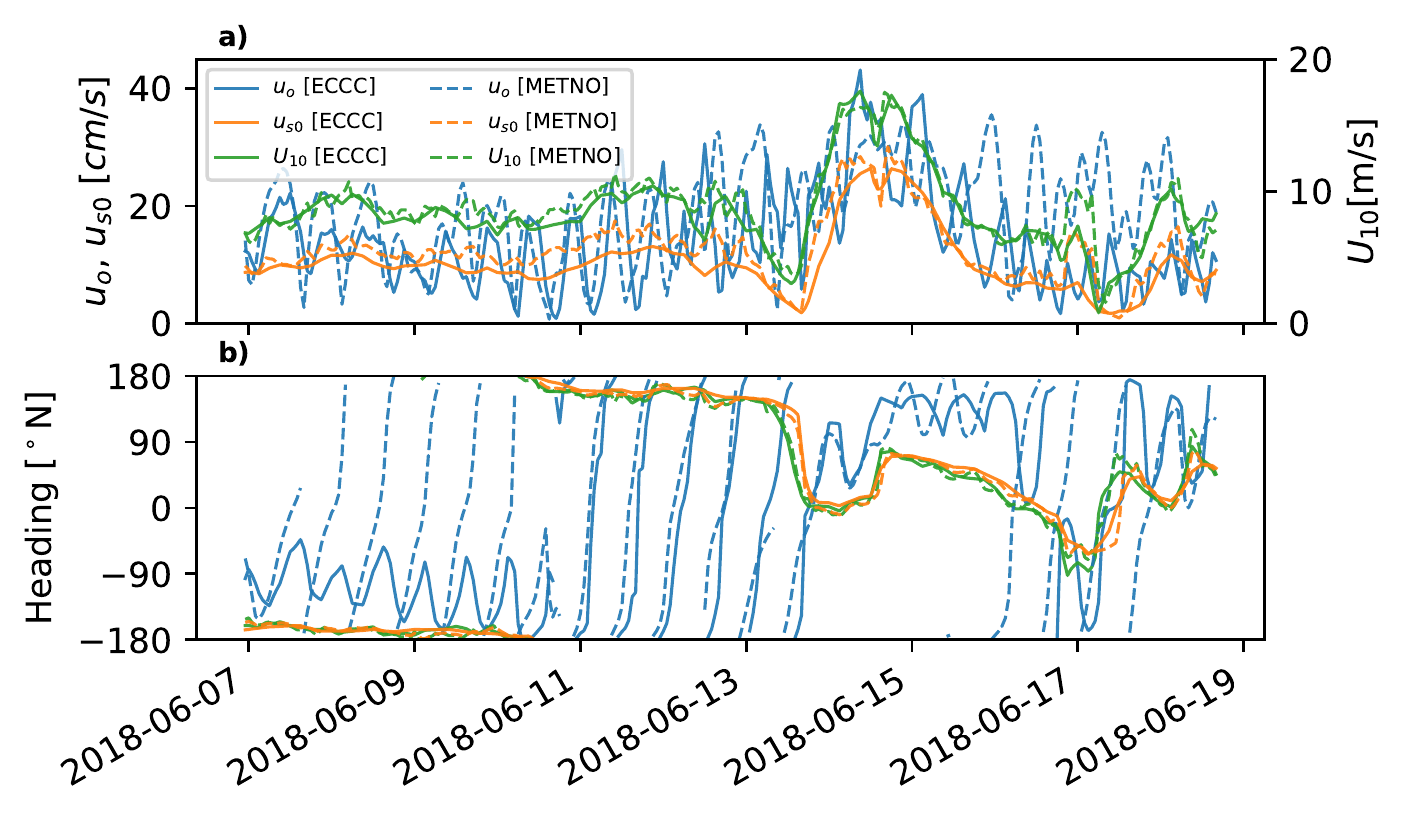}
   \caption{Forcing time series interpolated to the Osker drift track. \textbf{a)} shows the magnitude of the ocean current ($u_o$), surface Stokes drift ($u_{s0}$) and wind speed at 10 meters ($U_{10}$) for the ECCC prediction system (solid lines) and the METNO prediction system (dashed lines). \textbf{b)} shows the heading for the velocities in \textbf{a)} in degrees clockwise from north.}
   \label{fig:timeseries}
\end{figure}

\begin{figure}[ht]
   \noindent\includegraphics[width=\textwidth]{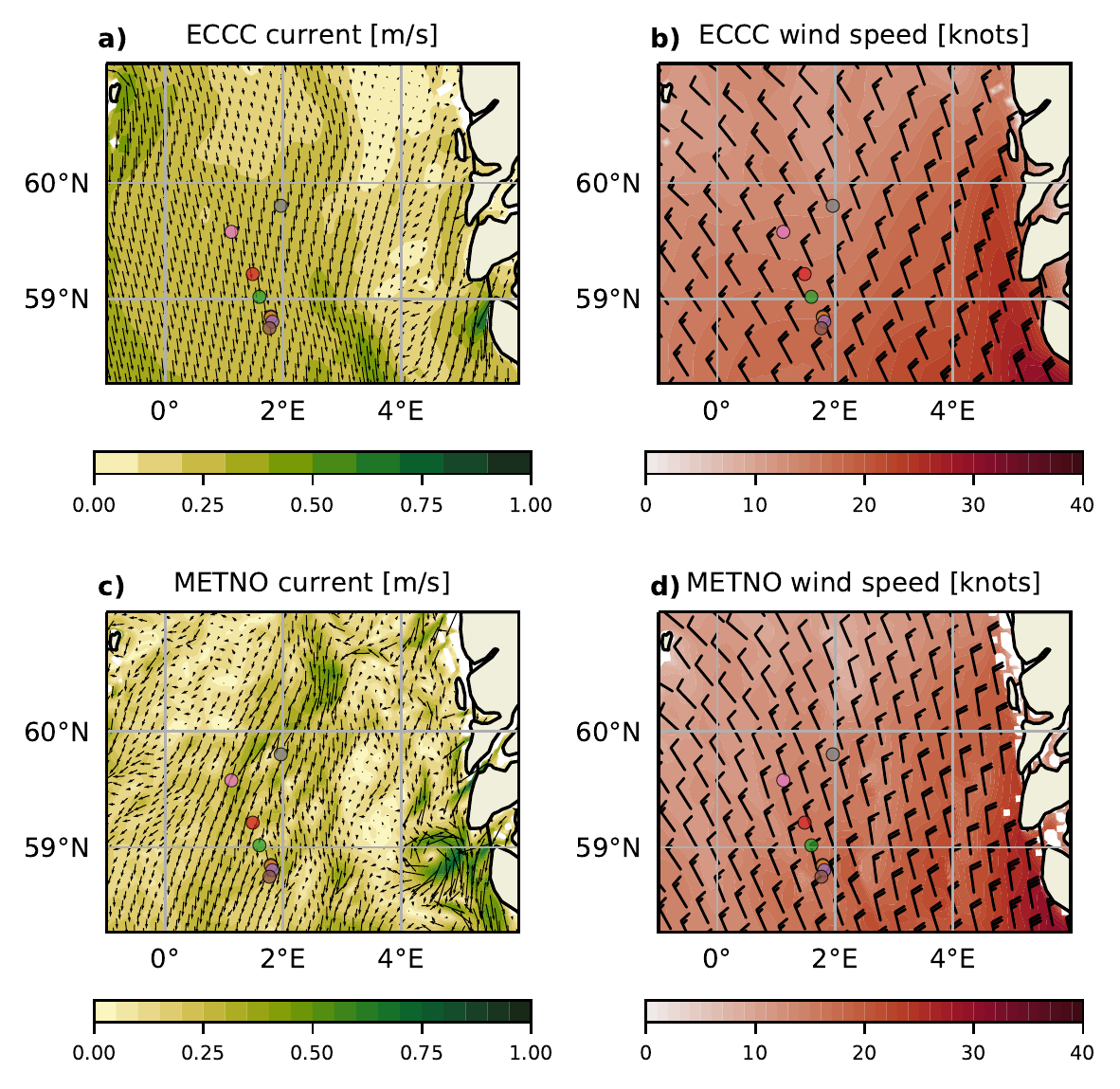}
   \caption{Snapshots of forcing fields on 13 June 2018 at 00Z. Surface velocities and 10 meter wind speeds are shown for the ECCC prediction system in panels \textbf{a)} and \textbf{b)} and for the METNO prediction system in panels \textbf{c)} and \textbf{d)}. Drifter locations at 13 June 2018 00Z are denoted by coloured dots and the drifter colour is shown in Figure~\ref{fig:map}.}
   \label{fig:forcing_map}
\end{figure}

The velocities produced by each the two prediction systems are linearly interpolated in space and time to each drifter track. An example for the Osker drifter track can be found in Figure~\ref{fig:timeseries}. There is little difference between the ECCC winds and METNO winds with both showing a peak wind speed of 18 m/s on 14 June 2018. The largest discrepancies between the ECCC and METNO prediction systems occurs between the ocean velocities. Figure~\ref{fig:forcing_map} shows snapshots of the surface currents and 10 m winds at 13 June 2018 showing the higher variability in the higher resolution METNO ocean velocities (Figure~\ref{fig:forcing_map}c) compared to the ECCC ocean velocities (Figure~\ref{fig:forcing_map}a). However, there do exist similarities in large scale features between the two ocean velocities as well as similarities in the 10 m wind field (Figures~\ref{fig:forcing_map}b and c for ECCC and METNO winds respectively). The surface Stokes drift (Figure~\ref{fig:timeseries}a) is also very similar between the ECCC and METNO models with the direction predominantly following the wind. Mean values for $u_{s0}/U_{10}$ for the ECCC and METNO prediction systems are 1.2\% and 1.4\% respectively. 

\subsection{Calculating $\alpha$ and $\beta$}

The implicit leeway coefficient $\alpha$ can be calculated directly from equation (\ref{eq:eulerian_leeway}) using the time series for each drifter of the modeled wind and ocean currents, i.e.
\begin{equation}\label{eq:alpha_temp}
   \alpha = \frac{\mathbf{u}_d - \mathbf{u}_o}{\mathbf{U}_{10}}.
\end{equation}
The ocean velocity $\mathbf{u}_o$ and 10 meter wind velocity $\mathbf{U}_{10}$ are interpolated in space and time to the location of the drifter. The interpolation is bilinear in space from the four adjacent grid points and linearly in time. The vector $\alpha$ is the leeway coefficient which produces an exact prediction at each time step. Therefore, the distribution of $\alpha$ will encompass all of the uncertainties in the model ocean and wind velocities as well as any uncertainties in the observed drift velocity. The velocity vectors are written in complex form so the real part is positive in the Eastward direction and the imaginary part is positive in the Northward direction. Therefore, the real part of $\alpha$ will be in the along-wind direction and the imaginary part in the cross-wind direction (negative to the right of the wind direction).

The same analysis can be applied to the explicit leeway model (\ref{eq:lagrangian_leeway}) to calculate $\beta$ where now the Stokes drift $u_{s}$ is explicitly included, i.e.
\begin{equation}\label{eq:beta_temp}
   \beta = \frac{\mathbf{u}_d - \mathbf{u}_s - \mathbf{u}_o}{\mathbf{U}_{10}}.
\end{equation}

Both equations (\ref{eq:alpha_temp}) and (\ref{eq:beta_temp}) are calculated for each recorded drifter position. This provides a time series of the complex value for $\alpha$ (or $\beta$), which can be used to estimate the mean and variance of the along-wind and cross-wind components. 

\section{Results}
\label{sec:results}

\begin{figure}[ht]
   \noindent\includegraphics[width=\textwidth]{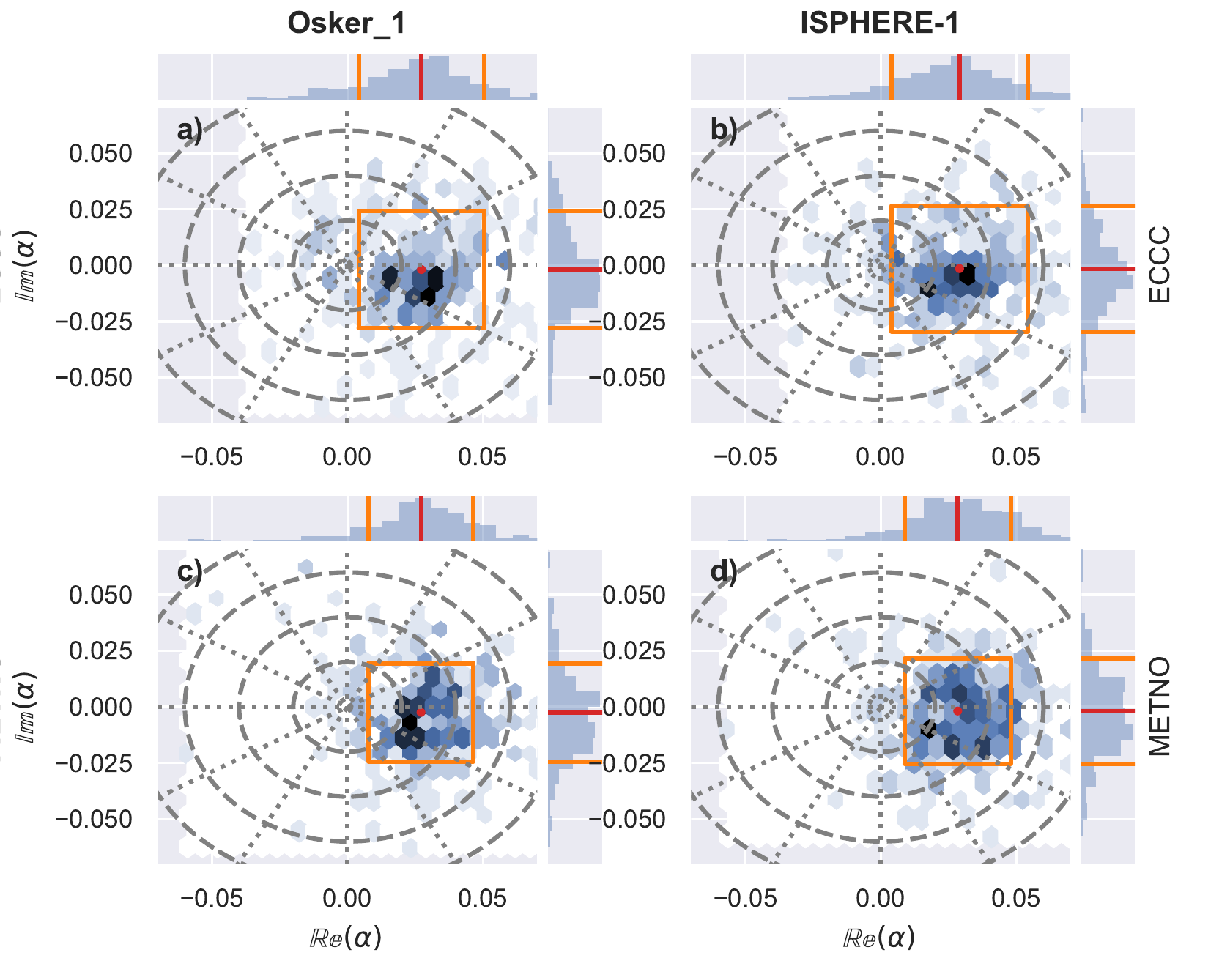}
   \caption{2D histograms of $\alpha$ (implicit Stokes drift) for two of the surface drifters and two different model forcing. \textbf{a)} and \textbf{b)} show the histograms for the ECCC forcing and \textbf{c)} and \textbf{d)} refer to the METNO forcing. }
   \label{fig:polar_surface_nostokes}
\end{figure}

The 2-D histogram of $\alpha$ for the surface drifters (Osker and iSPHERE) shows the mean and variance over the experiment period for both choices of environmental forcing (Figure~\ref{fig:polar_surface_nostokes}). Both the ECCC and METNO forcing yield a peak in the PDF at $|\alpha|$ of about 0.03 and an angle of less than $5^\circ$ to the right of the wind. The standard deviation of the along-wind and cross-wind components are between 0.02 and 0.03. For each 2-D histogram there is also the associated 1-D histogram of the along-wind component (above each 2-D histogram) and of the cross-wind component (to the right of each 2-D histogram). The mean value is shown in red, and the standard deviation of each component is shown in orange. The orange box encloses values which are within one standard deviation of the mean. The standard deviation is slightly higher using the ECCC forcing compared to the METNO forcing and both forcings give a standard deviation slightly greater in the cross-wind compared to the along-wind direction.

\begin{figure}[ht]
   \noindent\includegraphics[width=\textwidth]{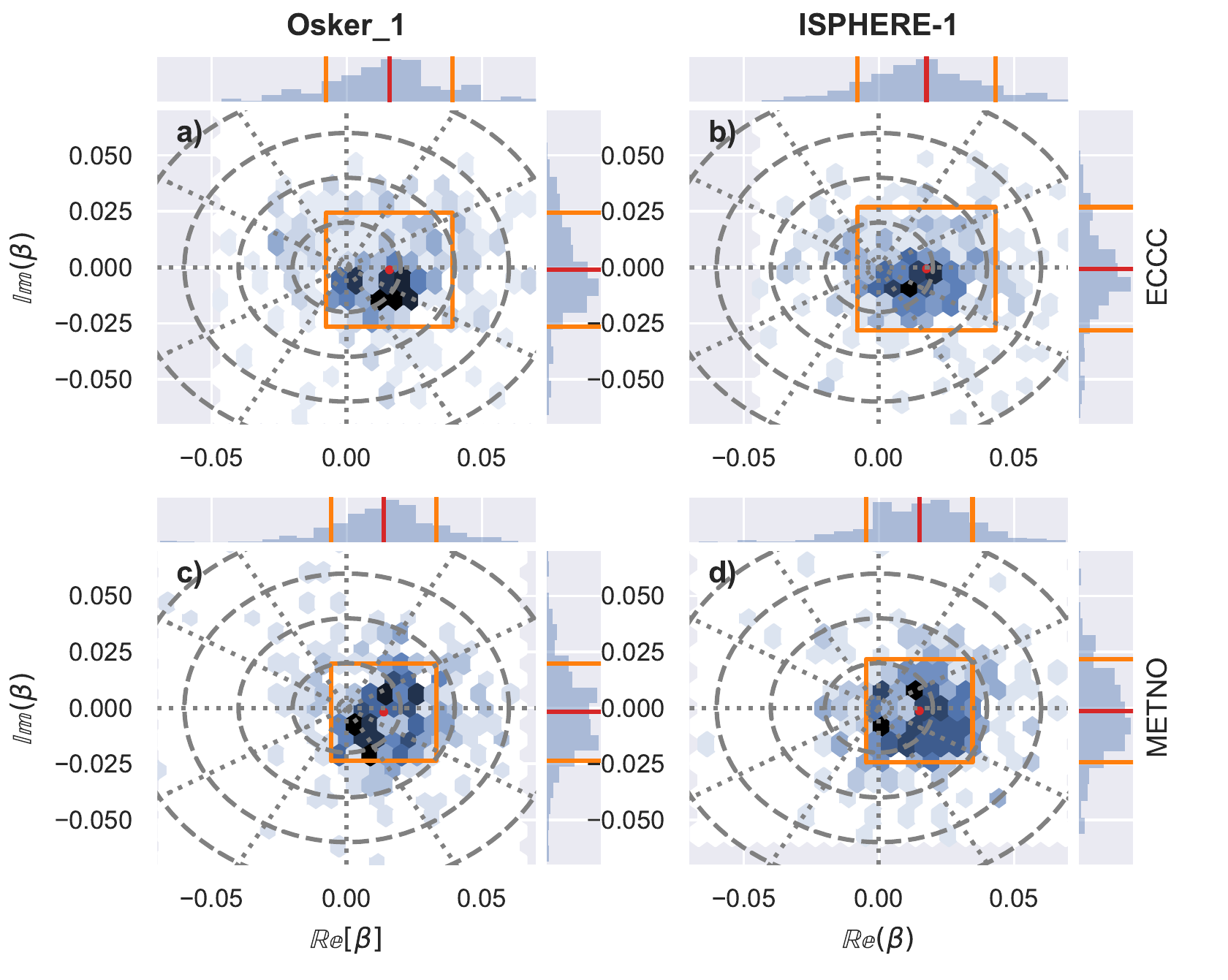}
   \caption{2D histograms of $\beta$ (explicit Stokes drift) for two of the surface drifters and two different model forcing. \textbf{a)} and \textbf{b)} show the histograms for the ECCC forcing and \textbf{c)} and \textbf{d)} refer to the METNO forcing. }
   \label{fig:polar_surface_stokes}
\end{figure}

The 2-D histogram for when the Stokes drift is explicitly included, as in (\ref{eq:beta_temp}), is shown in Figure~\ref{fig:polar_surface_stokes} and the results are shown in Table~\ref{tab:beta}. The along-wind component for $\beta$ is less when the Stokes drift is included, with a reduction of 0.011 and 0.013 compared to $\alpha$ for the ECCC and METNO forcing respectively. In addition, the standard deviation for $\beta$ is identical to that of $\alpha$ suggesting that including the Stokes drift has no appreciable effect on the accuracy of the prediction over the duration of the observations.

The same analysis is performed for the near-surface drifters (Roby and SCT) and the drogued drifters (CODE and SVP). Mean and standard deviation for the implicit and explicit Stokes drift are found in Table~\ref{tab:alpha} and Table~\ref{tab:beta} respectively. The 2-D histograms are not included here, but are available in the Supplementary Material.

\begin{table}[t]
   \caption{Mean with 95\% confidence interval along with the standard deviation of $\alpha$ for each drifter and choice of forcing. Mean and standard deviation are calculated from hourly values between 6 June 2018 and 18 June 2018 for a total of 283 measured values.}
   \label{tab:alpha}
   \begin{center}
      \begin{tabular}{ccccccc}
         \hline\hline
         Drifter & Effective Depth [m] & Forcing & mean (real) & mean (imag) & std dev (real) & std dev (imag) \\
         \hline
         Osker & 0.0 & ECCC & 0.027$\pm$0.003 & -0.002$\pm$0.004 & 0.023 & 0.026\\
         Osker & 0.0 & METNO & 0.027$\pm$0.003 & -0.002$\pm$0.003 & 0.019 & 0.022\\
         iSPHERE & 0.0 & ECCC & 0.029$\pm$0.003 & -0.001$\pm$0.004 & 0.025 & 0.028\\
         iSPHERE & 0.0 & METNO & 0.028$\pm$0.003 & -0.002$\pm$0.003 & 0.020 & 0.023\\
         Roby & 0.1 & ECCC & 0.023$\pm$0.003 & -0.002$\pm$0.03 & 0.024 & 0.019\\
         Roby & 0.1 & METNO & 0.019$\pm$0.002 & -0.005$\pm$0.002 & 0.015 & 0.017\\
         SCT & 0.2 & ECCC & 0.016$\pm$0.003 & -0.001$\pm$0.002 & 0.019 & 0.018\\
         SCT & 0.2 & METNO & 0.015$\pm$0.003 & -0.003$\pm$0.003 & 0.022 & 0.024\\
         CODE & 0.6 & ECCC & 0.009$\pm$0.003 & -0.008$\pm$0.004 & 0.025 & 0.026\\
         CODE & 0.6 & METNO & 0.009$\pm$0.003 & -0.008$\pm$0.003 & 0.022 & 0.021\\
         SVP & 15 & ECCC & 0.004$\pm$0.002 & -0.002$\pm$0.002 & 0.018 & 0.015\\
         SVP & 15 & METNO & -0.001$\pm$0.002 & 0.010$\pm$0.003 & 0.015 & 0.019\\
         \hline
      \end{tabular}
   \end{center}
\end{table}

\begin{table}[t]
   \caption{Mean with 95\% confidence intervals (CI) along with standard deviation of $\beta$ for each drifter and choice of forcing. Mean and standard deviation are calculated from hourly values between 6 June 2018 and 18 June 2018 for a total of 283 measured values.}
   \label{tab:beta}
   \begin{center}
      \begin{tabular}{ccccccc}
         \hline\hline
         Drifter & Effective Depth [m] & Forcing & mean$\pm$ 95\% CI (real) & mean$\pm$ 95\% CI (imag) & std dev (real) & std dev (imag) \\
         \hline
         Osker & 0.0  & ECCC & 0.016$\pm$0.003 & -0.001$\pm$0.003 & 0.023 & 0.025\\
         Osker & 0.0 & METNO & 0.014$\pm$0.003 & -0.002$\pm$0.003 & 0.019 & 0.022\\
         iSPHERE & 0.0 & ECCC & 0.018$\pm$0.003 & -0.001$\pm$0.004 & 0.026 & 0.028\\
         iSPHERE & 0.0 & METNO & 0.015$\pm$0.003 & -0.001$\pm$0.003 & 0.020 & 0.023\\
         Roby & 0.1 & ECCC & 0.012$\pm$0.003 & -0.001$\pm$0.003 & 0.025 & 0.019\\
         Roby & 0.1  & METNO & 0.007$\pm$0.002 & -0.005$\pm$0.002 & 0.016 & 0.017\\
         SCT & 0.2 & ECCC & 0.006$\pm$0.003 & 0.000$\pm$0.002 & 0.020 & 0.018\\
         SCT & 0.2 & METNO & 0.004$\pm$0.003 & -0.003$\pm$0.002 & 0.023 & 0.024`\\
         CODE & 0.6 & ECCC & 0.001$\pm$0.003 & -0.007$\pm$0.004 & 0.025 & 0.026\\
         CODE & 0.6 & METNO & 0.001$\pm$0.003 & -0.008$\pm$0.003 & 0.023 & 0.022\\
         SVP & 15 & ECCC & 0.003$\pm$0.002 & -0.002$\pm$0.002 & 0.018 & 0.015\\
         SVP & 15 & METNO & -0.001$\pm$0.002 & 0.010$\pm$0.003 & 0.015 & 0.019\\
         \hline
      \end{tabular}
   \end{center}
\end{table}

For the near-surface drifters the along-wind component of $\alpha$ and $\beta$ are both slightly less than for the surface drifters with a larger decrease observed with the SCT drifters than the Roby drifters. This is to be expected as the SCT drifters have a greater mean depth than the Roby drifters so the Stokes drift and the direct wind forcing on the drifter should be less than the surface drifters. In addition, the SCT drifters have three cylindrical rings attached to the base (Figure~\ref{fig:drifters}), which will act to increase the water-side drag. 

The drogued drifters have even smaller mean leeway coefficients than the other two types of drifters, with less variation between the implicit and explicit Stokes drift variation of leeway coefficient. However, explicitly including the Stokes drift for the shallower CODE drifter does reduce the required windage to a negligible value. 

The standard deviation, as can be seen from Table~\ref{tab:alpha} and Table~\ref{tab:beta},  does not vary greatly with drifter, choice of forcing or whether the Stokes drift is explicitly included or is implicit in the leeway coefficient. These facts suggest that the variability is most likely due to the ocean currents that are relatively uniform in the upper 15 m such as the barotropic tide and/or inertial currents. 

\begin{figure}[ht]
	\noindent\includegraphics[width=\textwidth]{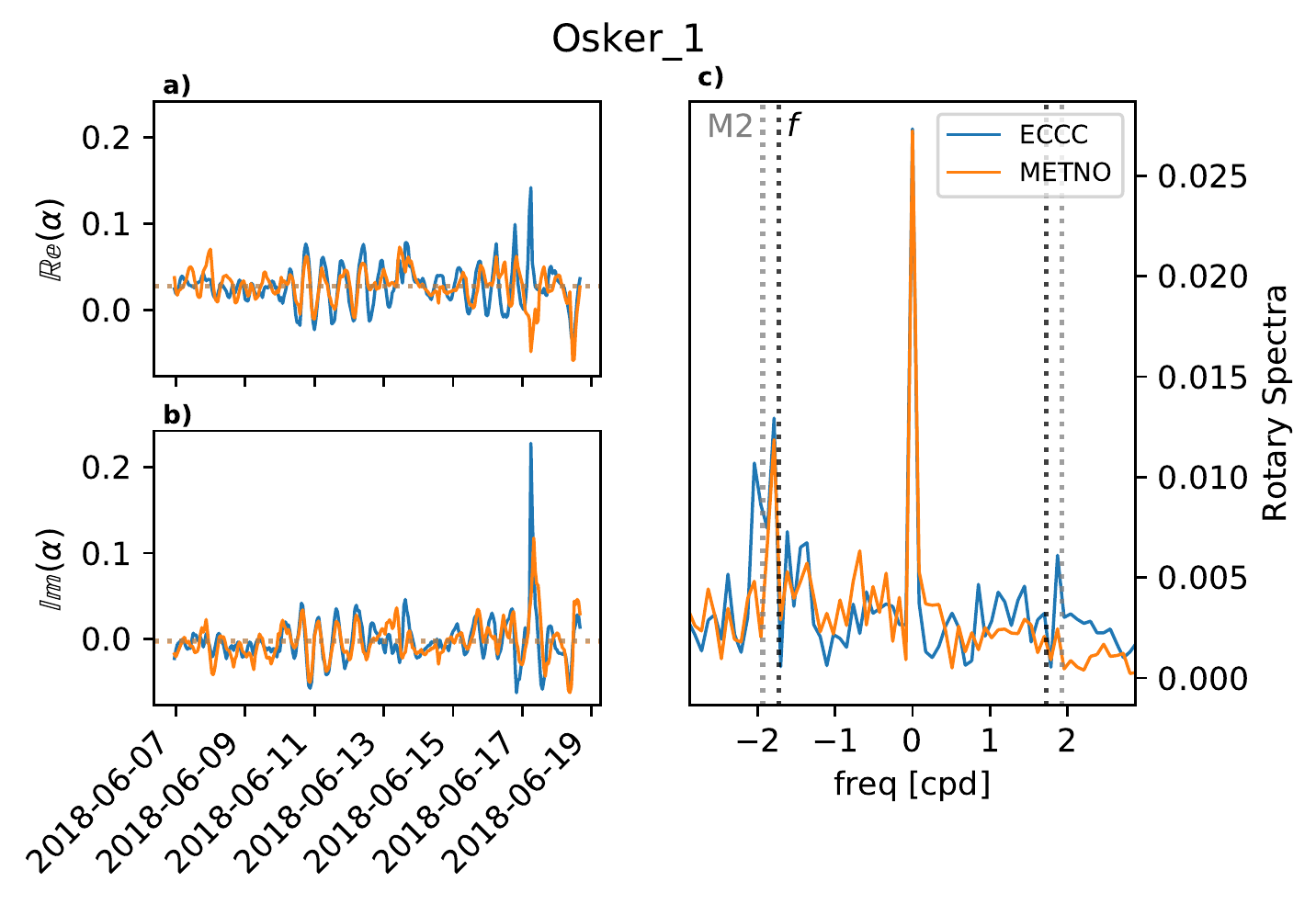}
   \caption{Time series of a) real part of $\alpha$, b) imaginary part of $\alpha$ and c) the magnitude of the rotary FFT of $\alpha$ for the Osker drifter. Horizontal dashed lines in a) and b) show the temporal mean values.  The vertical dashed lines in c) shows the inertial frequency (dark) and the M2 tidal frequency (light).}
   \label{fig:alpha_osker_nostokes}
\end{figure}

Another way to look at this ideal leeway coefficient is to investigate the time series to see if there is any observed periodicity. The time series for $\alpha$ of the Osker drifter calculated from (\ref{eq:alpha_temp}) where the real and imaginary part of $\alpha$ are shown in Figure~\ref{fig:alpha_osker_nostokes}a-b, respectively. Variability in both the real and imaginary part are consistent with inertial oscillations, which suggests that the leeway coefficient is compensating for errors in the amplitude and/or phase of the modeled inertial currents in order to reproduce the observed trajectory. The rotary FFT, which is the FFT of the complex valued $\alpha$, is shown in Figure~\ref{fig:alpha_osker_nostokes}c. The hourly sampling frequency over 12 days provides 283 samples providing a frequency resolution of 0.085 cpd. This is about 2.3 to 2.5 times the frequency difference between the M2 tide and the inertial frequency over the observed latitudes.

The spectrum in Figure~\ref{fig:alpha_osker_nostokes}c is dominated by the large mean response, which is similar in magnitude to the mean from the time series in Figure~\ref{fig:polar_surface_nostokes}, and by peaks at the inertial frequency for both choices of forcing. This inertial peak in $\alpha$ is nearly identical for all six drifters and both operational models (see Supplementary Material). For example, Figure~\ref{fig:alpha_svp_nostokes} shows the time series and FFT for the SVP drifter and the inertial peak is similar while the values at subinertial frequencies are generally less. This peak at the inertial frequency implies that the leading source of error for predictions are inertial currents. In both Figure~\ref{fig:alpha_osker_nostokes} and Figure~\ref{fig:alpha_svp_nostokes} there is a second peak near the M2 tidal frequency for the ECCC forcing, suggesting inaccuracies in the tidal component are also contributing to the velocity mismatch. 

\begin{figure}[ht]
	\noindent\includegraphics[width=\textwidth]{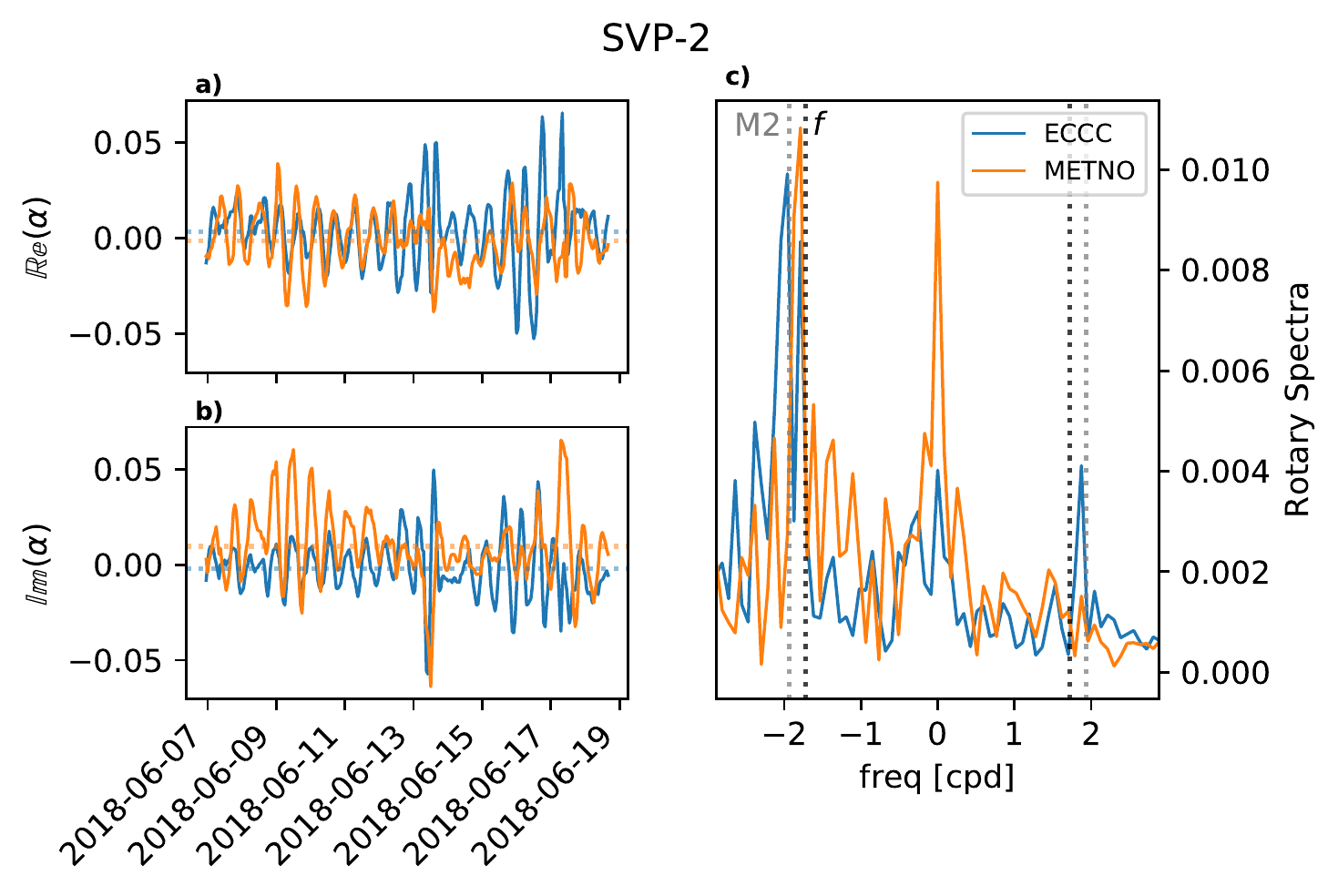}
   \caption{Time series of a) real part of $\alpha$, b) imaginary part of $\alpha$ and c) the magnitude of the rotary FFT of $\alpha$ for the SVP drifter. Horizontal dashed lines in a) and b) show the mean values and the vertical dashed lines in c) shows the inertial frequency.}
   \label{fig:alpha_svp_nostokes}
\end{figure}

\section{Discussion}
\label{sec:discussion}

It is inherently difficult to disentangle uncertainties associated with the marine environmental prediction systems and those associated with the leeway model in (\ref{eq:eulerian_leeway}) and/or (\ref{eq:lagrangian_leeway}). Using two operational prediction systems allows for the partial assessment of the forcing fields, especially when the two operational systems differ. However, the errors represent the sum of all the velocity errors, normalized by the wind speed, making determination of the largest source of error difficult. For the operational systems used in this study, ECCC and METNO, the statistics of the leeway coefficient, whether it be $\alpha$ or $\beta$, were very similar for the two models with differences being the least for the surface drifters and greatest for the drogued drifters. This result is consistent with~\citet{Dagestad_Rohrs_2019} who found that the surface drifter, specifically the iSPHERE, trajectories were strongly correlated with wind forcing and not strongly correlated with oceanic variability. As our two predictive systems provide similar wind forcing it appears that differences between the two systems are predominantly due to the oceanic component. 

While our results are similar to those of~\citet{Dagestad_Rohrs_2019} in that the surface drifters are much more sensitive to wind forcing than oceanic variability, the magnitude of the leeway coefficient are found to be smaller.  The leeway coefficient is found to be about 3\% when the Stokes drift is not included, and about 1.5\% when the Stokes drift is included, while~\citet{Dagestad_Rohrs_2019} found a value closer to 4\% when the Stokes drift was not included and 3\% when the Stokes drift is included. While it is not immediately obvious for the discrepancy, some of the reason may lie in the methodology, as we determine $\alpha$ (or $\beta$) from the entire trajectory while~\citet{Dagestad_Rohrs_2019} calculate $\alpha$ (or $\beta$) from several short forecasts of 48 hour duration. One hypothesis is that the use of shorter forecast lead times could limit the total accumulated uncertainty in the trajectory model. Also, allowing for the leeway coefficients $\alpha$ (or $\beta$) to be vectors provides a means (at least a posteriori) to correct for uncertainties in the ocean prediction system, such as errors due to the tides and inertial oscillations.

While the method has been presented here for drifting buoys, it could easily be used for any drifting object in the ocean. For example, the leeway coefficient for a ship adrift could be estimated with available data from the operational prediction system in order to improve future predictions. Such a method could also be dynamic as the leeway coefficient could be updated as more data become available. Uncertainty in high frequency motions in the ocean model, most notably from inertial currents, will undoubtedly influence estimates of the leeway coefficient over a short duration, but as long as data from a minimum of one inertial period is used to estimate the leeway coefficient then this uncertainty should have a minimal effect on the mean leeway coefficient.

The explicit inclusion of the Stokes drift, as calculated by a wave model, does not change the variability of the leeway coefficient and the magnitude of the Stokes drift, at least for surface objects, and can be approximated by a fraction of the wind speed. While it appears that a typical value of 1.3\%~\citep{Rascle_etal_2008} is appropriate for this case, there still exists variability in the literature on the range of 0.5 to 3\% depending on the sea state~\citep{Rascle_etal_2013}. Therefore, it is preferable to calculate the Stokes drift using available wave spectra, either from a numerical model or from observations, rather than a direct wind-based parameterization. Furthermore, for drifters in the upper meter but not right at the surface, it is important to calculate the e-folding depth and this is simplified by the use of an operational wave model.

\section{Conclusions}
\label{sec:conclusions}

Presented is an analysis of the leeway coefficient calculated for several drifter types using two different operational prediction systems. The leeway coefficient is calculated at each location for each drifter such that the prediction system exactly reproduces the observed trajectory. This method provides a timeseries of the leeway coefficient from which appropriate statistics can be calculated. In addition, calculating the leeway coefficient using this method provides the linear best estimate provided the forcing (currents, waves and winds) are unbiased. 

For the surface drifters, iSPHERE and the Osker, a mean wind induced drag of about 2.8$\pm$0.1\% and 1.6$\pm$0.2\% of the wind speed was found to reproduce the observed trajectories for the implicit and explicit leeway models respectively and did not depend on the choice of operational model. The standard deviation did not vary with the choice of leeway model and ranged between 2\% and 3\% for each of the along-wind and cross-wind components, with the cross-wind standard deviation being consistently larger than the along-wind standard deviation. This 3\% value for the wind drag is similar to previous reported values for the iSPHERE~\citep{Rohrs_Christensen_2015} and is the first study to look at the water following attributes for the Osker drifter. As the iSPHERE drifters are commonly used to track oil spills~\citep{Beloire_etal_2011}, our study suggests that Osker drifters should be equally well suited for this purpose. 

The two near-surface drifters, the Roby and SCT drifters which are both undrogued but with slightly deeper profiles in the water, had slightly reduced values for the leeway coefficient relative to the surface drifters. The Roby drifter had a similar value for $\alpha$ ($\beta$) of 2.3\% (1.2\%) for the ECCC forcing and 1.9\% (0.7\%) for the METNO forcing. The standard deviation between the two operational models is also slightly different with the ECCC values being 2.4\% to 2.5\% in the along-wind and 1.9\% in the cross-wind directions and the METNO values being 1.6\% in the along-wind and 1.7\% in the cross-wind directions. 

For the two drogued drifters, CODE and SVP, the leeway coefficient was found to be less than 1\%, which is consistent with results from previous studies~\citep{Niiler_etal_1995,Poulain_Gerin_2019}. The windage on the SVP should be negligible~\citep{Niiler_etal_1995} so these results most likely represent mean biases in the ocean model over these drifter tracks. By explicitly including the Stokes drift, the CODE drifters, with a mean depth of 0.6 m, have their along-wind leeway coefficient reduced from 0.9\% to 0.1\%. 

Investigations of the time series for the leeway coefficient show that the leading cause of variance in the leeway coefficient occurs at the inertial frequency, and that this appears to be consistent between surface drifters (e.g. OSKER) and 15 m drogued drifter (i.e. SVP). This suggests that the leeway coefficient is correcting for uncertainties in the ocean model inertial currents. As the drift in inertial oscillations is difficult to capture using leeway coefficients, ocean circulation models must be employed to resolve the magnitude, direction and phase of drift during the presence of inertial oscillations. However, ocean models will only capture an accurate response if the turbulence dynamics of the mixed layer are described well and the coupling to atmospheric models is implemented to a sophisticated degree, e.g. with small enough coupling time step~\citep{Christensen_etal_2018}.

%
\section*{datastatement}
Data is freely available from the corresponding author by request.

%
\section*{Acknowledgements}
Financial support for the drifter deployments was provided through the Government of Canada's Oceans Protection Plan. The field work was carried out during the 2018 Oil on water campaign with generous assistance from NOFO, the Norwegian Clean Seas Association for Operating Companies. KHC, KFD, JR and {\O}B gratefully acknowledge the financial support of the Research Council of Norway through the CIRFA project, grant no 237906.

%



%
%


%

\newpage
\section*{Supplementary material}

\begin{figure}[ht]
	\noindent\includegraphics[width=\textwidth]{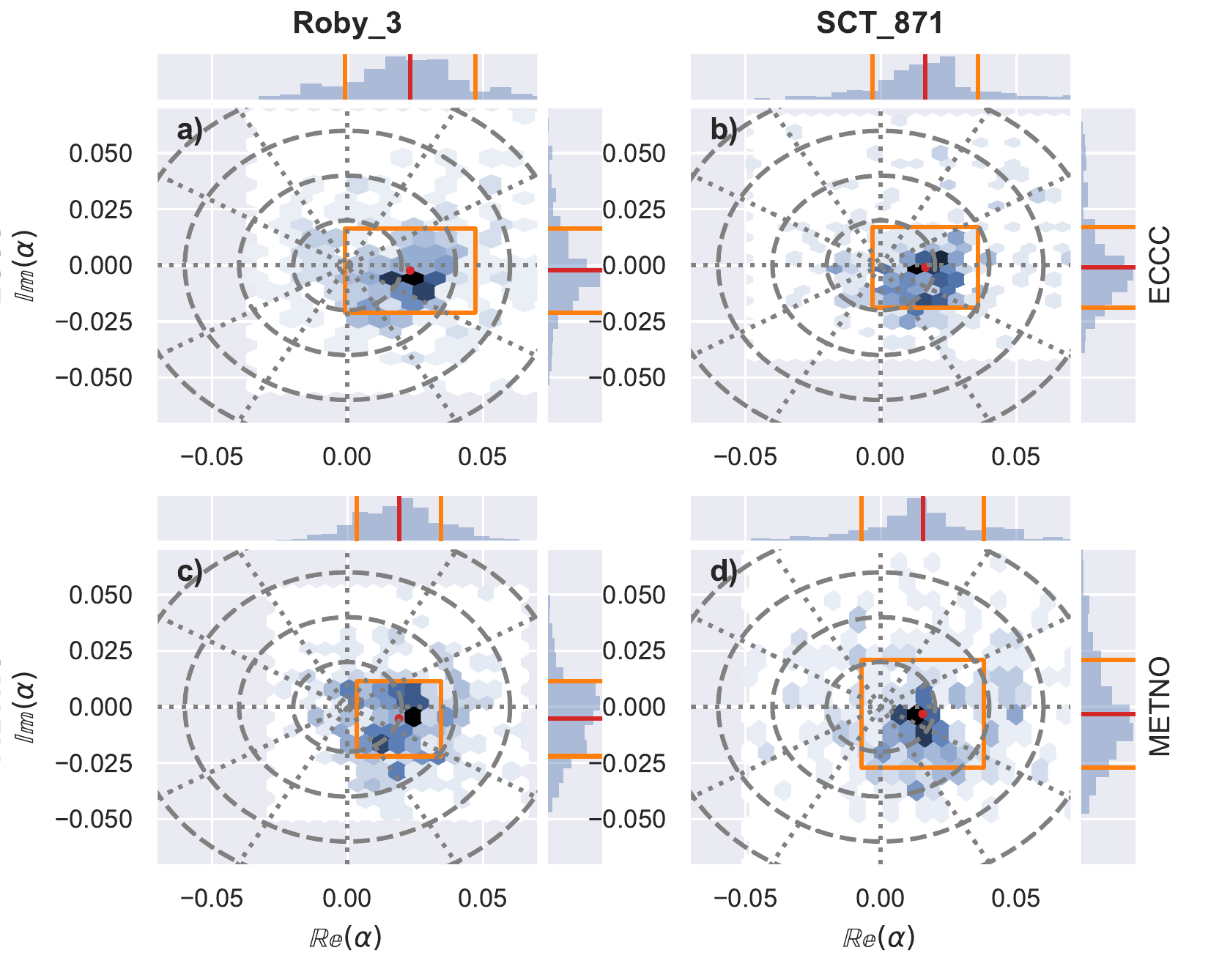}
   \caption{2D histograms of $\alpha$ (implicit Stokes drift) for the two near-surface drifters and two different model forcing. \textbf{a)} and \textbf{b)} show the histograms for the ECCC forcing and \textbf{c)} and \textbf{d)} refer to the METNO forcing. }
   \label{fig:polar_int_nostokes}
\end{figure}

\begin{figure}[ht]
	\noindent\includegraphics[width=\textwidth]{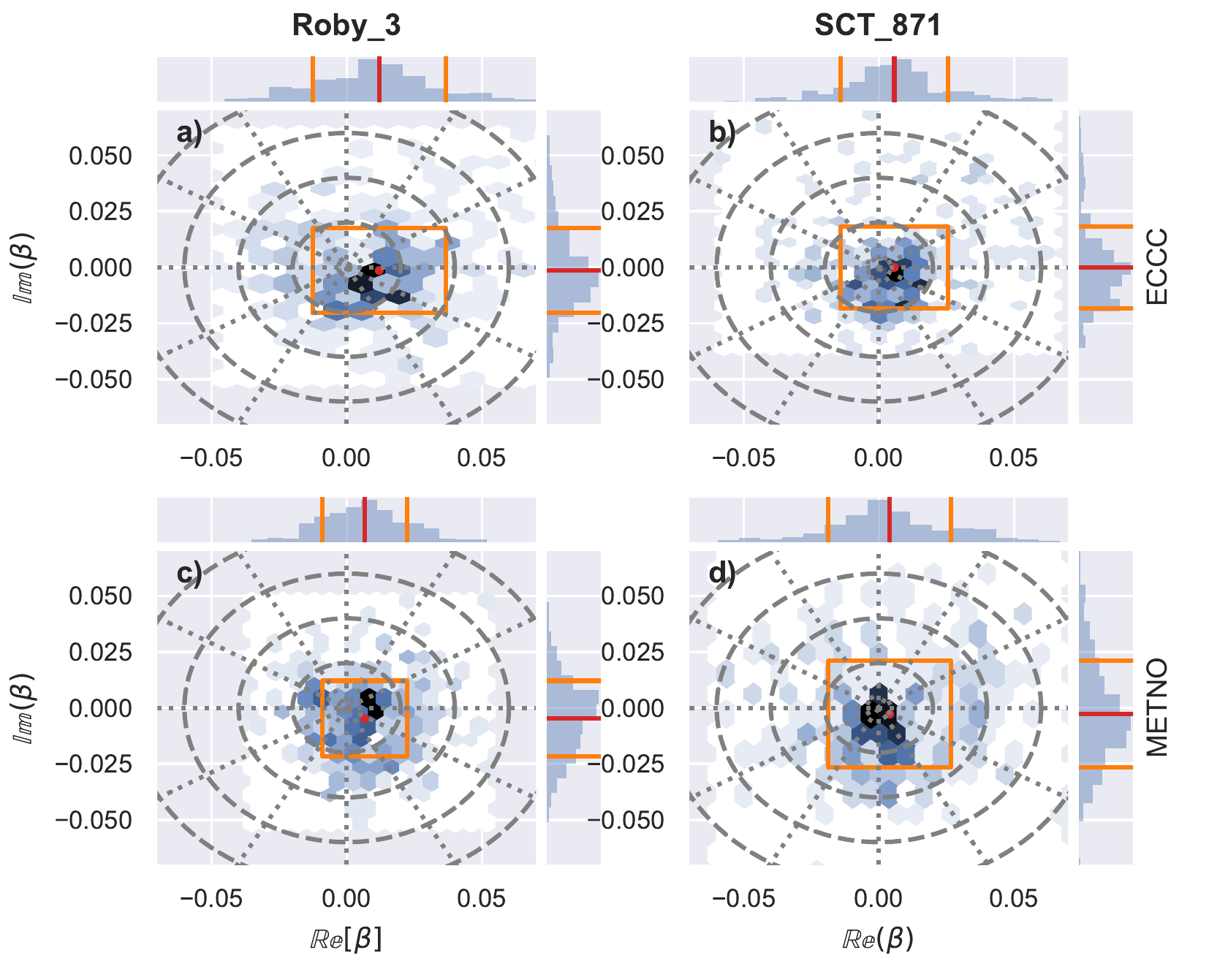}
   \caption{2D histograms of $\beta$ (explicit Stokes drift) for the two near-surface drifters and two different model forcing. \textbf{a)} and \textbf{b)} show the histograms for the ECCC forcing and \textbf{c)} and \textbf{d)} refer to the METNO forcing. }
   \label{fig:polar_int_stokes}
\end{figure}

\begin{figure}[ht]
	\noindent\includegraphics[width=\textwidth]{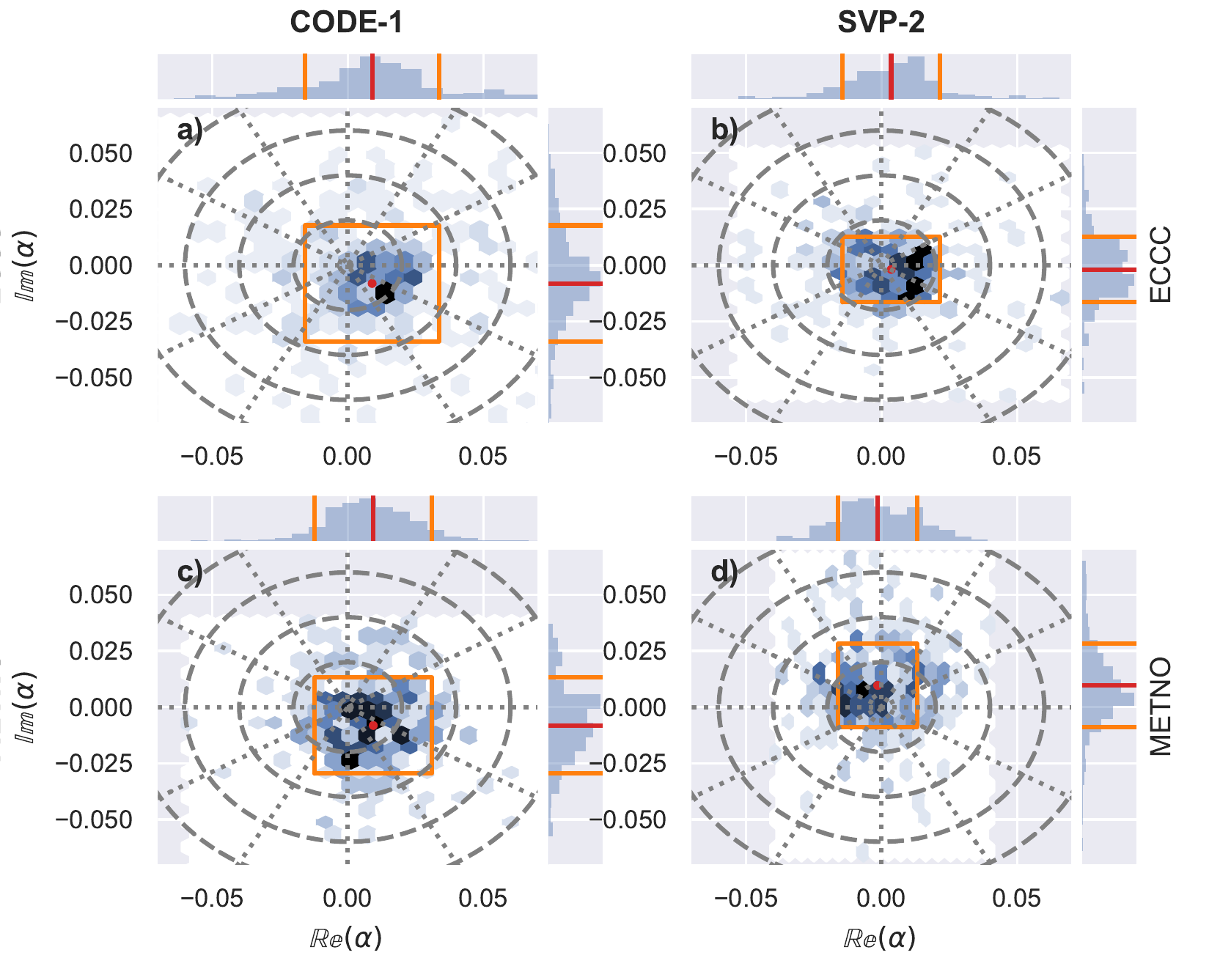}
   \caption{2D histograms of $\alpha$ (implicit Stokes drift) for the two drogued drifters and two different model forcing. \textbf{a)} and \textbf{b)} show the histograms for the ECCC forcing and \textbf{c)} and \textbf{d)} refer to the METNO forcing. }
   \label{fig:polar_drogue_nostokes}
\end{figure}

\begin{figure}[ht]
	\noindent\includegraphics[width=\textwidth]{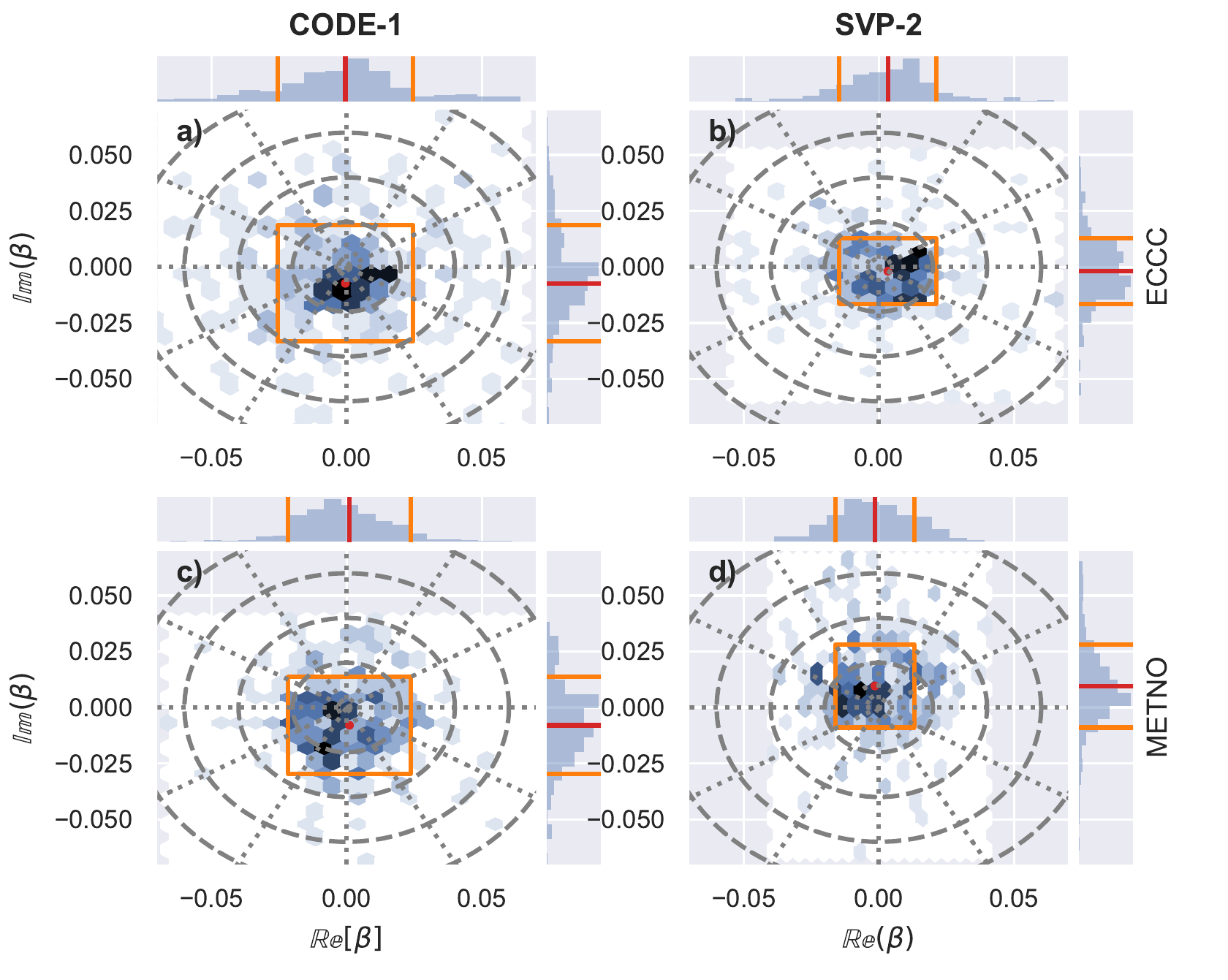}
   \caption{2D histograms of $\beta$ (explicit Stokes drift) for the two drogued drifters and two different model forcing. \textbf{a)} and \textbf{b)} show the histograms for the ECCC forcing and \textbf{c)} and \textbf{d)} refer to the METNO forcing. }
   \label{fig:polar_drogue_stokes}
\end{figure}

Figures~\ref{fig:polar_int_nostokes}-\ref{fig:polar_drogue_stokes} shows the two-dimensional histograms of the leeway coefficient for both the implicit and explicit Stokes drift and the two operational prediction systems used in the study. These are shown for the near-surface and drogued drifters.

\begin{figure}[ht]
   \noindent\includegraphics[width=\textwidth]{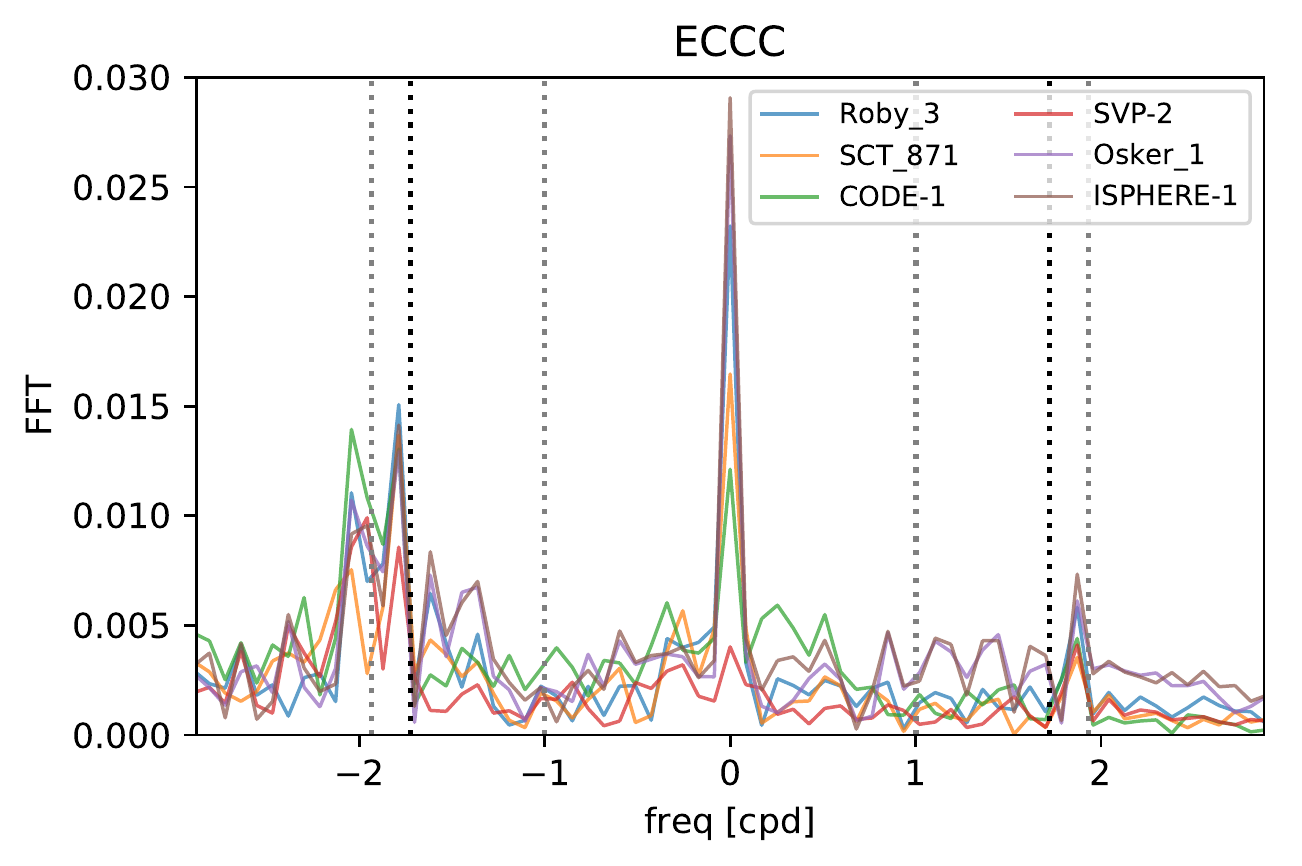}
   \caption{Spectra of $\alpha$ for ECCC forcing.}
   \label{fig:alpha_eccc}
\end{figure}

\begin{figure}[ht]
   \noindent\includegraphics[width=\textwidth]{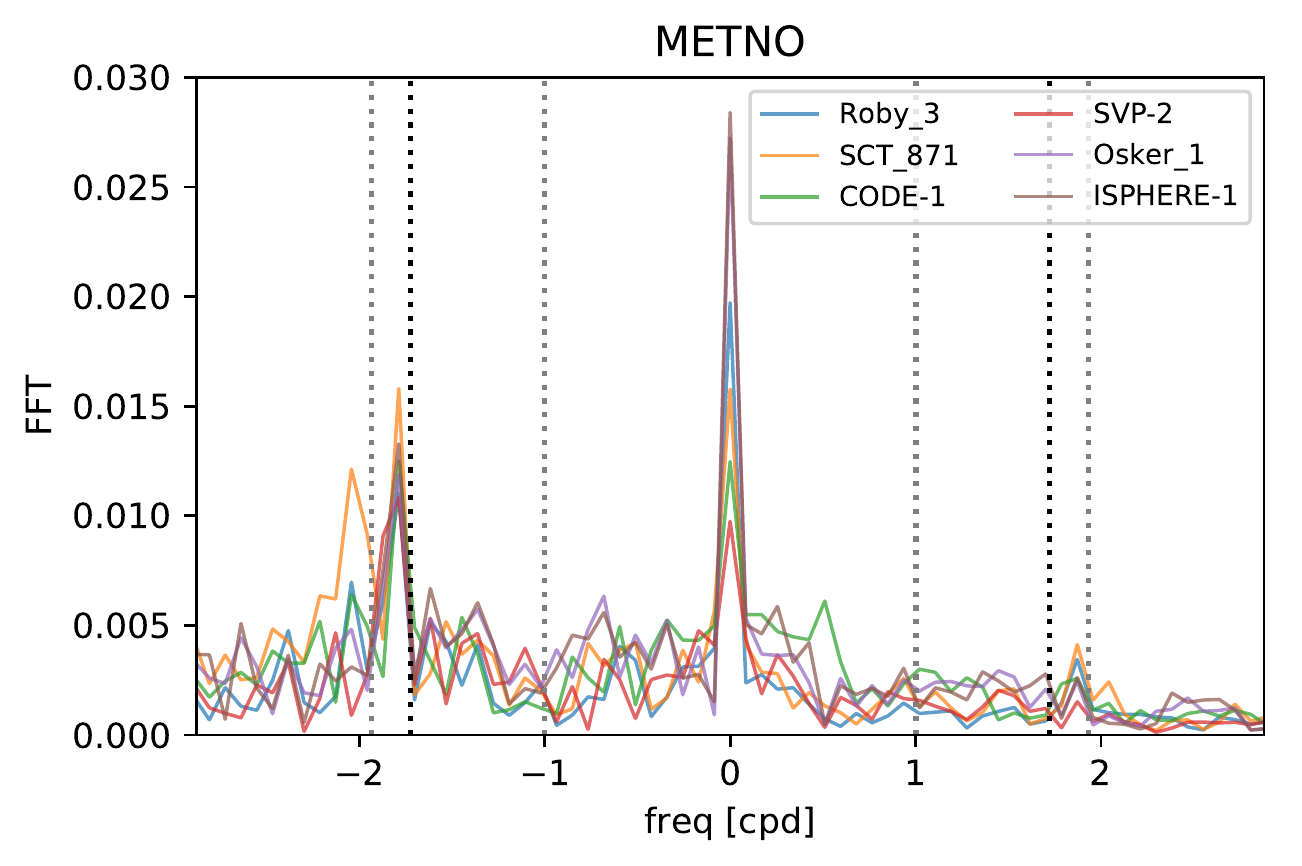}
   \caption{Spectra of $\alpha$ for METNO forcing.}
   \label{fig:alpha_metno}
\end{figure}

\begin{figure}[ht]
   \noindent\includegraphics[width=\textwidth]{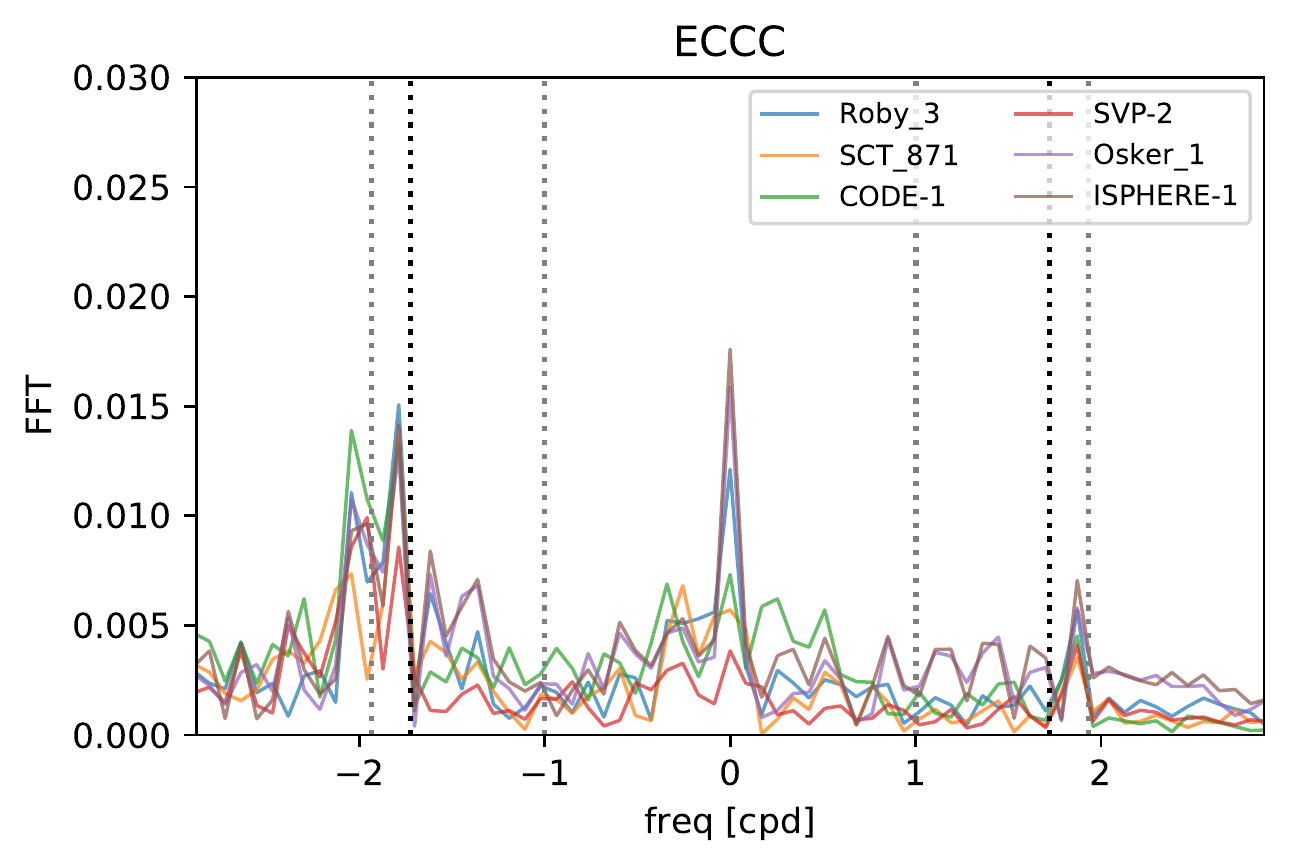}
   \caption{Spectra of $\beta$ for ECCC forcing. Vertical dashed lines show the M2 tidal frequency, inertial frequency and K1 tidal frequency from highest frequency to lowest respectively.}
   \label{fig:beta_eccc}
\end{figure}

\begin{figure}[ht]
   \noindent\includegraphics[width=\textwidth]{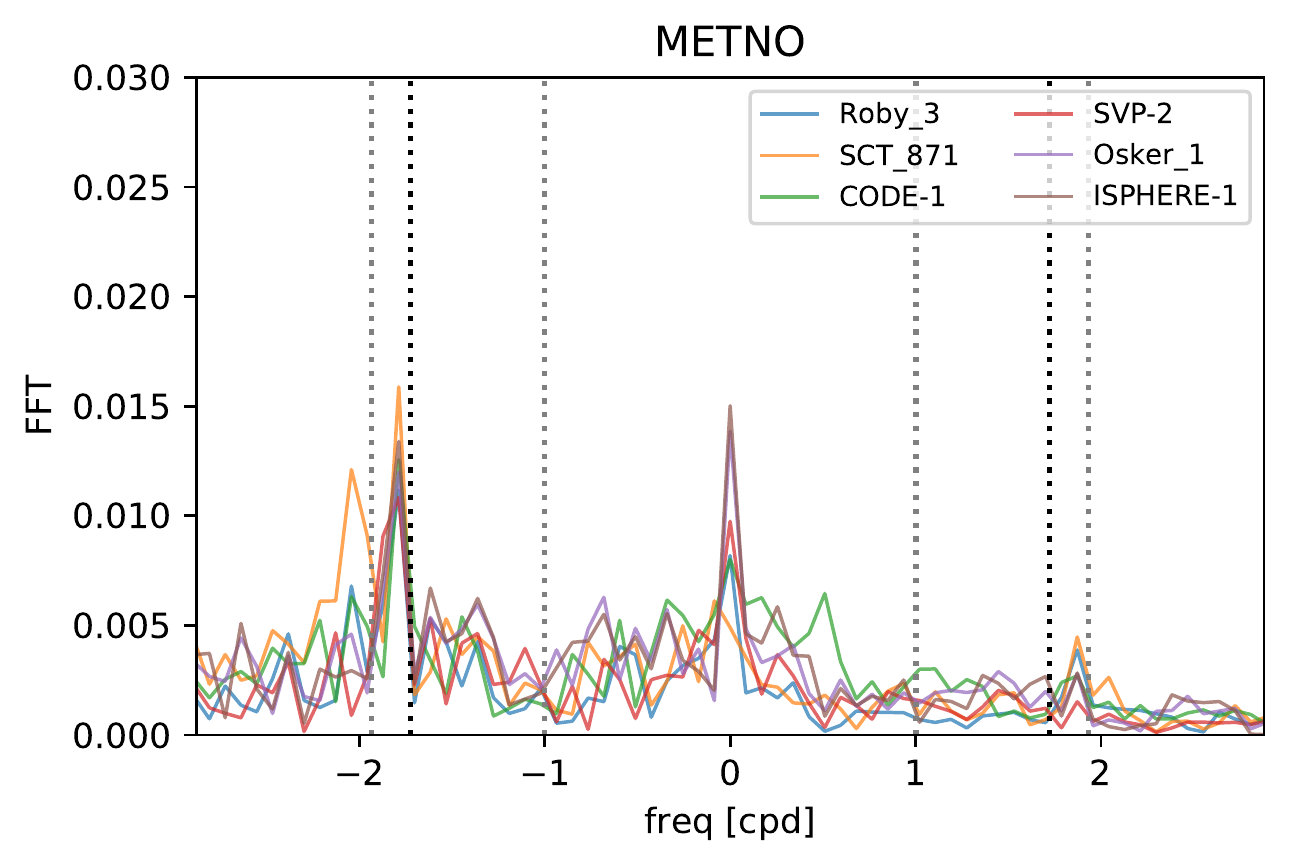}
   \caption{Spectra of $\beta$ for METNO forcing. Vertical dashed lines show the M2 tidal frequency, inertial frequency and K1 tidal frequency from highest frequency to lowest respectively.}
   \label{fig:beta_metno}
\end{figure}

Figures~\ref{fig:alpha_eccc}-\ref{fig:beta_metno} show the spectral plots for $\alpha$ and $\beta$ for each of the different forcings.

\newpage
\section*{Supplementary material}

\end{document}